\documentclass[sigconf]{aamas2023/aamas}

\setcopyright{ifaamas}
\acmConference[AAMAS '23]{Proc.\@ of the 22nd International Conference
on Autonomous Agents and Multiagent Systems (AAMAS 2023)}{May 29 -- June 2, 2023}
{London, United Kingdom}{A.~Ricci, W.~Yeoh, N.~Agmon, B.~An (eds.)}
\copyrightyear{2023}
\acmYear{2023}
\acmDOI{}
\acmPrice{}
\acmISBN{}

\acmSubmissionID{503}

\title{Mediated Multi-Agent Reinforcement Learning}

\author{Dmitry Ivanov}
\affiliation{
  \institution{HSE University}
  \city{Saint Petersburg}
  \country{Russia}}
\affiliation{
  \institution{Technion}
  \city{Haifa}
  \country{Israel}}
\email{divanov@campus.technion.ac.il}

\author{Ilya Zisman}
\affiliation{
  \institution{HSE University}
  \city{Saint Petersburg}
  \country{Russia}}
\email{iazisman@edu.hse.ru}

\author{Kirill Chernyshev}
\affiliation{
  \institution{HSE University}
  \city{Saint Petersburg}
  \country{Russia}}
\email{chernyshevk1212@gmail.com}

\begin{abstract}
  The majority of Multi-Agent Reinforcement Learning (MARL) literature equates the cooperation of self-interested agents in mixed environments to the problem of social welfare maximization, allowing agents to arbitrarily share rewards and private information. This results in agents that forgo their individual goals in favour of social good, which can potentially be exploited by selfish defectors. We argue that cooperation also requires agents' identities and boundaries to be respected by making sure that the emergent behaviour is an equilibrium, i.e., a convention that no agent can deviate from and receive higher individual payoffs. Inspired by advances in mechanism design, we propose to solve the problem of cooperation, defined as finding socially beneficial equilibrium, by using mediators. A mediator is a benevolent entity that may act on behalf of agents, but only for the agents that agree to it. We show how a mediator can be trained alongside agents with policy gradient to maximize social welfare subject to constraints that encourage agents to cooperate through the mediator. Our experiments in matrix and iterative games highlight the potential power of applying mediators in MARL.\footnote{This is a slightly updated version of this \href{https://dl.acm.org/doi/abs/10.5555/3545946.3598618}{[publication]}. Please cite the published version. The code is available \href{https://github.com/dimonenka/mediatedMARL}{[here]}.}
\end{abstract}

\usepackage[utf8]{inputenc} 
\usepackage[T1]{fontenc}    
\usepackage{hyperref}       
\usepackage{url}            
\usepackage{booktabs}       
\usepackage{amsfonts}       
\usepackage{nicefrac}       
\usepackage{microtype}      
\usepackage{xcolor}         
\usepackage{multirow}

\usepackage{mathtools}
\usepackage{caption}
\usepackage{subcaption}
\usepackage{amsfonts}
\usepackage{mathrsfs}
\usepackage{balance} 
\usepackage{graphicx}


\begin{document}

\pagestyle{fancy}
\fancyhead{}

\maketitle

\section{Introduction}\label{sec:introduction}

The cooperation of self-interested agents is an elusive concept to define and measure, especially in temporally and spatially extended environments typical for Multi-Agent Reinforcement Learning (MARL). In these environments, agents may have multiple low-level actions available that may not be inherently cooperative or competitive. The aggregated effect of these actions affects the rewards of all agents and the state of the environment in ways that may not be easy to disentangle. This is further complicated by games lasting for multiple turns, over which the effect of agents' actions accumulates. An approach alluring in its simplicity is to measure cooperation as social welfare, i.e., some aggregate (usually, the sum total) of cumulative rewards of all agents \cite{perolat2017multi}. This allows complete freedom in the choice of training procedures, including arbitrary reward manipulations, information sharing, and parameter sharing (examples are provided in Section \ref{sec:introduction_related}).

In this paper, we challenge this view. While maximizing social welfare is a relevant problem of its own, not every solution should count as a solution to cooperation. Unlike fully cooperative MARL settings where all agents share a common objective, mixed (i.e., general-sum) settings imply self-interested agents with clear boundaries. Blending their interests and blurring their boundaries does not make the agents more cooperative -- it makes the setting itself more cooperative. Instead, cooperation is only meaningful if it is a consequence of the rational decision-making of strategic agents that act in their own best interests. This is only possible if the agents act in an \textit{equilibrium}, i.e., a convention that none of the agents can deviate from and increase their individual reward. Defining cooperation as converging to socially beneficial equilibrium (like tit-for-tat) is sometimes referred to as conditional cooperation, as opposed to unconditional cooperation that focuses on maximizing social welfare \cite{zhao2022proximal}. As we discuss later, very few existing approaches address conditional cooperation.

\begin{table*}[t]
\centering
\caption{The effect of the mediator on prisoner's dilemma. The mediator acts on behalf of all agents that agree to it, i.e., commit. In prisoner's dilemma (a), the dominant action is to defect. In the mediated prisoner's dilemma (b), the dominant action is to commit if the mediator adopts the following strategy: cooperate if both agents committed, otherwise defect.}
\label{table:pd}
\begin{tabular} { cc }
(a) Prisoner's Dilemma & (b) Mediated Prisoner's Dilemma \\
\begin{tabular}{@{}lcc@{}}
    \toprule
              & Defect & Cooperate \\
    Defect    & 0, 0   & 2, 2     \\
    Cooperate & 2, 2  & 4, 4      \\ \bottomrule
\end{tabular} &
\begin{tabular}{@{}lccc@{}}
    \toprule
              & Defect & Cooperate & Commit \\
    Defect    & 0, 0   & 7, -5  & 0, 0    \\  
    Cooperate & -5, 7  & 2, 2  &  -5, 7   \\  
    Commit & 0, 0  & 7, -5  &  2, 2   \\ \bottomrule
\end{tabular}
\end{tabular}
\end{table*}

Designing generic MARL algorithms that promote or create socially beneficial equilibria is a complex and largely unsolved problem. Prospective solutions can come from the field of mechanism design (and related fields, such as information design and contract design) \cite{parkes2015economic}. This field studies how to implement trusted entities known as mechanisms that interact with self-interested agents in ways that both consider their incentives and achieve desirable social outcomes. Taking auction design as an example, the agents' incentives can be valuations over an item and a desirable social outcome can be allocating the item to the agent with the highest valuation. From the economic perspective, unconditional cooperation is unrealistic since one cannot arbitrarily modify the rewards or incentives of an economic agent the way one may be able with an artificial agent. In our example, the designer cannot simply ask agents to disclose their valuations and give away the item for free to the highest number because the agents would have no incentives to report truthfully. An example more relevant to MARL is self-driving vehicles \cite{conitzer2019designing}: while we can encode the rewards of the vehicles, we cannot encode the rewards of the people that use them. Instead, desirable social outcomes should be aligned with agents' incentives -- a property known as Incentive-Compatibility (IC). A vast literature on designing IC mechanisms is waiting to be adapted to mixed MARL. This direction has recently gotten attention in several papers that we discuss in Section \ref{sec:introduction_related}.

In this paper, we show how conditional cooperation in MARL can be solved via mediators \cite{monderer2009strong}. A mediator is a benevolent entity that can act on behalf of the agents that agree to it. In a mediated augmentation of a game, each agent first decides whether act on its own or to let the mediator act for it, an action that we call `commit'. We call the set of committed agents a `coalition'. Then, the original game is played between the mediator acting for the coalition and the rest of the agents. Note that the mediator can have a unique policy for each coalition. Crucially, an agent always has the opportunity to refuse to commit (serving as an opportunity to misreport), e.g., if it finds the mediator incapable of achieving an acceptable reward or if it wants to exploit other agents cooperating through the mediator. The potential impact of mediators is illustrated in Table \ref{table:pd} on the example of the prisoner's dilemma.

We are interested in applying mediators to complex sequential games where optimal policies are unknown in advance and have to be learned. Specifically, we propose to train both agents and the mediator jointly with MARL. This creates challenges atypical for unconditional cooperation: not only the mediator has to learn a policy that maximizes social welfare (for each coalition), but it also has to encourage agents to commit, i.e., ensure compatibility with their incentives. We show how to formulate this as a constrained optimization problem, as well as how to solve this problem using the method of Lagrange multipliers and dual gradient descent.

Additionally, we introduce the frequency of agents' commitment as a hyperparameter that we denote as commitment window $k$. When $k=1$, agents' commitment lasts for a single time step. When $k>1$, agents are only allowed to commit each $k$-th time step and the commitment lasts for $k$ time steps (i.e., the commitment is more committing). As a result, the mediator has a higher potential to maximize social welfare for larger $k$.

There are multiple advantages to our approach. First, any emergent behaviour is by design an equilibrium because agents act in their own best interests (i.e., maximize their own rewards) and can deviate if they find it beneficial. Second, \citet{monderer2009strong} show that in symmetric games there always exists a mediator with a socially optimal strategy such that committing is an equilibrium for the agents. This means that there is no loss in achievable social welfare when using mediators compared to unconditionally maximizing social welfare. For asymmetric games, we empirically demonstrate the power of mediators. Third, the mediator's strategy is fair in the sense that each committed agent receives the expected payoffs that are at least as high as if it acted on its own. This is unlike the existing techniques that artificially encode fairness into a centralized objective \cite{jiang2019learning,zimmer2021learning}. Fourth, the mediator only promotes cooperation between the committed agents and deters the free-riding of the rest. These are the properties of reciprocity that naturally emerge from the Incentive-Compatibility of our solution, which is unlike the existing attempts at training reciprocal agents based on arbitrary reward sharing and heuristics \cite{lerer2017maintaining,peysakhovich2018consequentialist,eccles2019learning,baker2020emergent}. Fifth, the rate of commitment to the mediator presents a generic measure of cooperation.

We experimentally validate our procedure to train agents with the mediator in several variants of prisoner's dilemma \cite{rapoport1965prisoner} and public good game \cite{bergstrom1986private}, including one-shot and sequential games. For each game, we first analyze the. We find that naively training the mediator to maximize social welfare may result in agents refusing to commit, either due to misalignment of their and the mediator's interests or due to them trying to exploit other committed agents. We show that this can be addressed by considering agents' incentives through additional constraints when training the mediator. Additionally, we investigate the effect of varying the commitment window $k$ and find that higher $k$ gives the mediator more power.

\subsection{Related Work}\label{sec:introduction_related}

\citet{zhao2022proximal} distinguish two kinds of cooperation in MARL settings. Unconditional cooperation refers to cooperation independent of what the opponents are doing. Reciprocity-based cooperation refers to cooperation \textit{iff} others cooperate, e.g., tit-for-tat in the iterated prisoner's dilemma.

We use a similar but more general classification of unconditional and conditional cooperation that better reflects the literature. We define the problem of unconditional cooperation as the maximization of social welfare. The problem of conditional cooperation additionally has a condition that no agent has incentives to deviate, i.e., that the agents are in equilibrium.

\paragraph{Unconditional cooperation.}

The majority of unconditional cooperation techniques are based on modifying or replacing agents' rewards with `intrinsic' preferences \cite{cittern2015reinforcement,lerer2017maintaining,peysakhovich2018consequentialist,hughes2018inequity,peysakhovich2018prosocial,eccles2019learning,wang2019evolving,jiang2019learning,baker2020emergent,durugkar2020balancing,yang2020learning,baumann2020adaptive,zimmer2021learning,ivanov2021balancing,phan2022emergent}. The intrinsic preferences can be either rewards of other agents or rewards learned by agents or a third party to guide agents to social welfare maximizing outcomes. The most direct approach is to train each agent to directly optimize social welfare, but this is susceptible to issues like free-riding and credit assignment that can be addressed by exploiting reward decomposition available in mixed environments. These techniques hardwire other-regarding preferences into agents and are typically not concerned with equilibria from the perspective of maximizing the original rewards.

The few alternative techniques typically require parameter or information sharing. An example of the former is the parameterization of all agents with identical neural networks \cite{gupta2017cooperative}. This technique sidesteps the conflict of interests by hardwiring reciprocity into agents: trying to exploit automatically reflects. An example of the latter is the social influence that uses a communication channel to maximize impact on message recipients \cite{jaques2019social}. This technique ignores the potential incentives of agents to manipulate each other through communication channels.

While these techniques are typically framed as solutions to cooperation, the problem they solve is more akin to the fully-decentralized MARL \cite{zhang2018fully,yang2020cm3,konan2022iterated}. In this setting, agents can have varying reward functions, but the collective goal is to maximize the globally averaged return over all agents, i.e., the social welfare. The agent-specific reward functions serve as an instrument to address credit assignment, so it indeed makes sense to not be concerned with equilibria. In contrast, mixed MARL implies selfish interests, and while social welfare is undoubtedly a useful performance measure, treating its unconditional maximization as a unanimously shared goal is a misleading shortcut.

\paragraph{Conditional cooperation.}

Some approaches to conditional cooperation look for existing equilibria with high social welfare. Learning with Opponent-Learning Awareness (LOLA) and its modifications \cite{foerster2018learning,willi2022cola,zhao2022proximal} leverage alternative gradient updates that shape the opponent's learning to guide it to cooperative equilibria. As a result, it can learn reciprocal strategies, e.g., tit-for-tat in the repeated prisoner's dilemma. LOLA has multiple limitations: it is only applicable to two-player games, requires access to the transition dynamics, and assumes that the opponent learns using first-order gradient-based methods. Furthermore, LOLA updates require read access to the opponent's parameters. While this can be circumvented by learning the opponent's model based on their behaviour, it comes at the expense of performance.

Other approaches change the rules of the game such that self-interested agents prefer to cooperate, i.e., such that outcomes with high social welfare become equilibria. At this point, the line between unconditional and conditional cooperation may become blurry. After all, modifying an agent's rewards with intrinsic preferences could be reinterpreted as paying the agent extrinsically by a third party. However, once additional rewards are considered extrinsic payments, a natural question is what is the minimal payment scheme such that agents still cooperate. None of the papers described in the unconditional cooperation section ask this question.

The question of optimal payments is central to the economic field of contract design \cite{grossman1992analysis}. Adapting contract design to MARL is attempted in the concurrent work of \citet{contracts2023marl}. This work proposes for one of the agents to take the role of the principal that may pay other agents. However, their empirical algorithm assumes that the payment condition is pre-determined and only learns the payment amount. Designing a generic algorithm that can learn optimal payment schemes (both conditions and amounts) by either one of the agents or a third party is an open problem.

Other solutions to conditional cooperation can be considered reward redistribution \cite{ibrahim2020reward}, which can also be framed as taxation \cite{zheng2020ai,zheng2022ai}; and similarity-based cooperation \cite{oesterheld2022similarity}, which is an extension of program equilibrium \cite{rubinstein1998modeling,tennenholtz2004program}.

In recent years, there has been a rising interest in communication under competition in MARL \cite{noukhovitch2021emergent,blumenkamp2021emergence}. In contrast to communication in fully cooperative environments, mixed environments may incentivize self-interested agents to manipulate others through their messages, preventing reliable and mutually beneficial communication from being established. A principled way to resolve this issue could potentially come from the field of Bayesian Persuasion \cite{kamenica2011bayesian}. There have been extensions of Bayesian Persuasion to online multi-receiver settings \cite{castiglioni2021multi}, as well as to MARL \cite{lin2023information}.


\section{Problem Setup}\label{sec:setup}

\subsection{Markov Games}\label{sec:setup_markov_games}

Markov game is a standard formalization of spatially and temporally extended environments typical for MARL \cite{littman1994markov}. It is defined as a tuple $\mathcal{G} = (S, N, (\mathcal{A}_i)_{i \in N}, (\mathcal{O}_i)_{i \in N}, T, (r_i)_{i \in N})$. Let $S$ be the set of all possible states $s$, $N$ be the set of agents, $\mathcal{A}_i$ be the set of actions $a_i$ available to the agent $i$ in all states. Let $\mathcal{O}_i: S \rightarrow O_i$ be the observation function, where $O_i$ is the set of observations $o_i$ of agent $i$. Let $T: S \times (\mathcal{A}_i)_{i \in N} \rightarrow \Delta(S)$ be the transition function, where $\Delta$ denotes a set of discrete probability distributions. This function specifies the effect of the agents' actions on the state of the environment. We enumerate the sequences of sampled transitions with time-steps $t$. Let $r_i: S \times (\mathcal{A}_j)_{j \in N} \rightarrow \mathcal{P}(\mathcal{R})$ be the reward function for each agent $i$, where $\mathcal{P}$ is a set of continuous probability distributions, $\mathcal{R} \subseteq \mathbb{R}$. Let $\tilde{r}_{i, t} \sim r_i(s_t, \textbf{a}_t)$ denote sampled rewards.

Let $\tilde{R}_{i, t} = \overset{\infty}{\underset{l=t}{\sum}} \left[ \gamma^{l-t} \tilde{r}_{i, l}  \right]$ be the discounted cumulative reward a.k.a. the return, where $\gamma \in [0, 1)$ denotes the discount factor. Let $\pi_i: O_i \rightarrow \Delta(\mathcal{A}_i)$ be the policy of agent $i$. Let $V_{i}(o_i) = \mathop{\mathbb{E}}_{\boldsymbol\pi, r_i, T, p(s \mid \boldsymbol\pi, \mathcal{O}_i(s) = o_i)}[\tilde{R}_i]$ be the value function. The agent seeks the policy $\pi_i$ that maximizes the value $V_i$ in each observation.

\subsection{Mediators}\label{sec:setup_mediators} 

A \textit{mediator} can be viewed as an additional entity that may act in the game on behalf of a subset of agents. The agents interact with the mediator by optionally sending it messages, and the mediator acts for those agents that sent it a message. Crucially, an agent may refrain from sending a message and act independently from the mediator. Formally, \citet{monderer2009strong} define mediator as a tuple $\mathcal{M}=((M_i)_{i \in N}, \textbf{c} = (\textbf{c}_C)_{\emptyset \neq C \subseteq N})$, where $M_i$ is a finite set of messages that agent $i$ may send to the mediator, $C$ is a subset of agents that sent messages to the mediator referred to as \textit{coalition}, and $\textbf{c}_C: \textbf{M}_C \rightarrow \Delta((\mathcal{A}_i)_{i \in C})$ is the correlated strategy (the joint policy) for the coalition $C$. Each agent has a utility function over the outcomes, the expectation of which it rationally maximizes.

A special case that we focus on is the \textit{minimal} mediator. A mediator is called minimal if each message space $M_i$ is a singleton, meaning that agents' interaction with the mediator is limited to agreeing to enter the coalition. We refer to this action as \textit{committing}. To uniquely define a minimal mediator, specifying $M_i$ becomes unnecessary. A strategy of the mediator that makes a unanimous commitment an equilibrium is called mediated equilibrium. Crucially, \citet{monderer2009strong} show that any mediated equilibrium can be implemented by a minimal mediator, as well as that mediated equilibrium that maximizes social welfare always exists in symmetric games.

Note that an agent cannot misreport its commitment to the mediator, i.e., deviate while pretending to commit. Instead, the action of refusing to commit itself serves as misreporting. A weaker variant of a mediator that only recommends actions is also explored in the economic literature \cite{rogers2014asymptotically,kearns2014mechanism,kearns2015robust,cummings2015privacy}. Applying this idea to MARL could be an interesting new direction.

It is also important to note that \citet{monderer2009strong} primarily explore mediators through the lens of strong mediated equilibria, i.e., equilibria robust to deviations of groups of agents. Since the number of different groups grows exponentially with the number of agents, finding strong mediated equilibria with RL is a challenging problem. We leave it to future work.

\subsection{Markov Mediators}\label{sec:setup_markov_mediators} 

The approach of \citet{monderer2009strong} implies fixed strategies of the mediator. Instead, we treat the mediator as a separate agent with its own goals and train it with RL alongside other agents. To this end, we introduce Markov mediators.

Let $\mathcal{M}=((\mathcal{O}_C^M)_{\emptyset \neq C \subseteq N}, (r_C^M)_{\emptyset \neq C \subseteq N}, k)$ be minimal Markov mediator. Let $\mathcal{O}_C^M: S \rightarrow O_C^M$ be the mediator's observation function for coalition $C$, where $O_C^M$ is the set of the mediator's observations. As the observations, we will use a tuple $o_C^M = ((o_i)_{i \in C}, C)$. Note that alternative choices of the mediator's observations like $o_C^M = ((o_i)_{i \in N}, C)$ or $o_C^M = (s, C)$ create an asymmetry of information between the mediator and the agents, which may serve as an additional incentive for the agents to commit. However, this would require access to additional information during execution.

Let $r_C^M: S \times (\mathcal{A}_i)_{i \in N} \rightarrow \mathcal{P}(\mathcal{R})$ be the mediator's reward function for coalition $C$. We are only concerned with mediators with the goal of increasing the utilitarian social welfare of the agents. For this reason, as the mediator's reward we will use the sum of rewards of the agents in the coalition: $\tilde{r}_{C, t}^M = \sum_{i \in C} \tilde{r}_{i, t}$. Let $k \in \mathbb{Z}^+$ be the commitment window. Each $k$ steps of the game, an agent may either commit to the mediator or choose to play independently for the next $k$ steps. Agents can only commit when $t$ is divisible by $k$.

We define the policy $\boldsymbol\pi_C^M: O_C^M \rightarrow \Delta((\mathcal{A}_i)_{i \in C})$, the return $\tilde{R}_{C, t}^M = \overset{\infty}{\underset{l=t}{\sum}} \left[ \gamma^{l-t} \tilde{r}_{C, l}^M  \right] = \sum_{i \in C} \tilde{R}_{i, t}$, and the value $V_C^M(o_C^M) = \mathop{\mathbb{E}}[\tilde{R}_C^M] = \sum_{i \in C} \mathbb{E} [\tilde{R}_{i}]$ of the mediator for each coalition $C$ similarly to those of the agents. Notice that the return and the value of the mediator decompose into the respective sums over the coalition.




\section{Algorithm}\label{sec:algorithm}

We now discuss how to train the Markov mediator with RL. We first describe our practical implementations of agents and the mediator, including neural architectures and loss functions, and then dive deeper into potential objectives for the mediator. Note that we write all expressions as expectations, but in practice, these are approximated as empirical averages over sampled transitions.

\begin{figure*}[t]
    \centering
    \includegraphics[width=\linewidth]{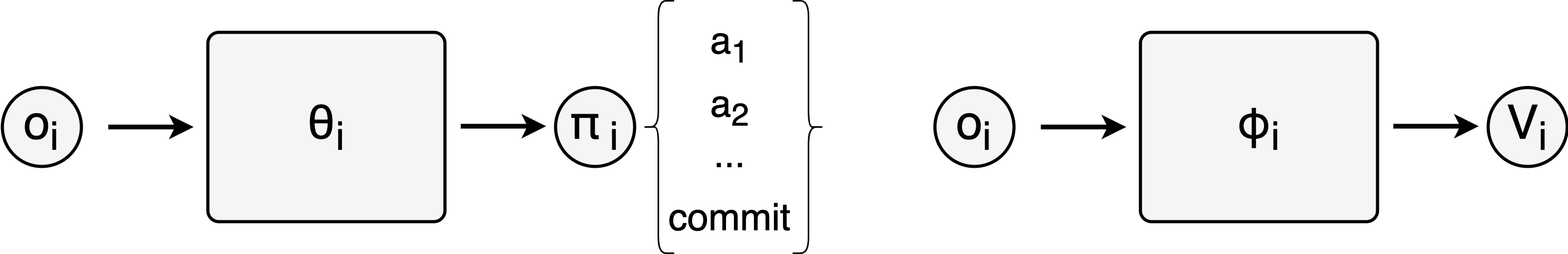}
    \caption{Schematic illustration of our architectures of the actor (left) and the critic (right) of the agents, $k=1$.}
    \label{fig:architecture_agents}
\end{figure*}

\begin{figure*}[t]
    \centering
    \includegraphics[width=\linewidth]{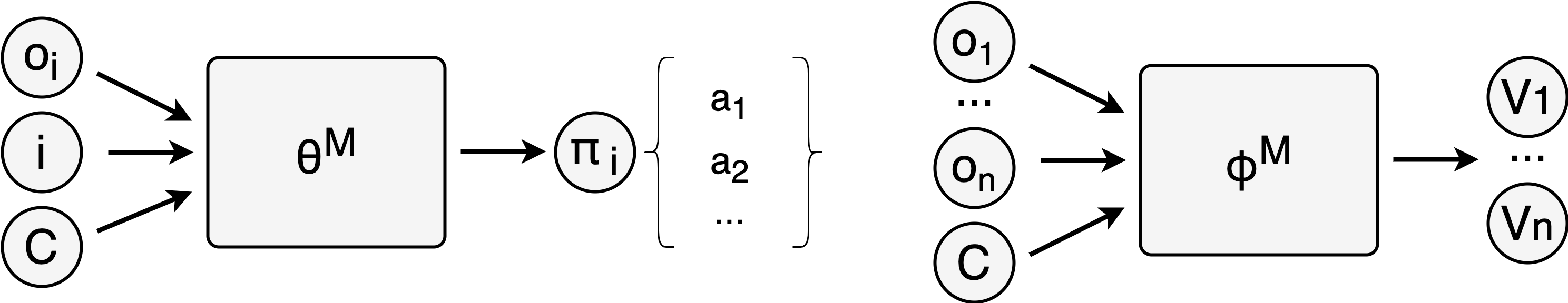}
    \caption{Schematic illustration of our architectures of the actor (left) and the critic (right) of the mediator, $k=1$.}
    \label{fig:architecture_mediator}
\end{figure*}

\subsection{Practical Implementations}\label{sec:algorithm_practical}

Both the agents and the mediator are trained via Actor-Critic frameworks \cite{konda2000actor,sutton2018reinforcement}. The actor represents the policy $\pi(o)$, whereas the critic represents an approximation of the value function $\tilde{V}(o)$. Both actor and critic can be parameterized with neural networks \cite{A3C}. The architectures are illustrated in Figures \ref{fig:architecture_agents} and \ref{fig:architecture_mediator} and are described below.

\paragraph{Agents}

The agents are trained independently: an agent $i$ has its own actor $\pi_{\theta_i}(o_i)$ and critic $\tilde{V}_{\phi_i}(o_i)$, respectively parameterized by $\theta_i$ and $\phi_i$, that are trained only on its own experience. The respective loss functions $L$ are the negated policy gradient for the actor and the squared temporal difference for the critic:

\begin{equation}
\label{loss_agent_actor}
L(\theta_i) = - \mathbb{E} (\tilde{r}_{i, t} + \gamma \tilde{V}_{\phi_i}(o_{i, t+1}) - \tilde{V}_{\phi_i}(o_{i, t})) \log \pi_{\theta_i}(o_{i, t})
\end{equation}

\begin{equation}
\label{loss_agent_critic}
L(\phi_i) = \mathbb{E} (\tilde{r}_{i, t} + \gamma \tilde{V}_{\phi_i}(o_{i, t+1}) - \tilde{V}_{\phi_i}(o_{i, t}))^2
\end{equation}

MARL literature routinely leverages parameter and experience sharing for both actor \cite{gupta2017cooperative} and critic \cite{lowe2017multi}, but we opt out of these techniques to ensure that agents are selfish and individual, and therefore that any cooperation observed is rational rather than a consequence of hardwired reciprocity (as discussed earlier).

We apply minimal changes to adapt agents to the presence of the mediator. When $k=1$, the only change is that each agent's actor is augmented with an additional action, i.e., to commit to the mediator. The effect of this action is that the mediator takes control over the agent for the current time-step.

When $k > 1$, the mediator acts for the agent for $k$ steps at a time, and the commitment action is only available to the agents every $k$ time-steps. This necessitates several additional changes. First, the current time-step $t$ is concatenated to an agent's observation to let it know whether it can commit. Second, in the succeeding $k-1$ states after an agent commits, it effectively acts off-policy and therefore is not trained. Third, committing for $k$ steps effectively transitions the agent from $s_t$ directly to $s_{t+k}$, and in the process yields the total discounted reward of $\sum_{l=t}^{t+k-1} \gamma^{l-t} \tilde{r}_{i, l}$. For this reason, (only) when an agent commits, the 1-step temporal difference in the loss functions (\ref{loss_agent_actor}) and (\ref{loss_agent_critic}) is replaced with the k-step temporal difference: $(\tilde{r}_{i, t} + \gamma \tilde{V}_{i, t+1} - \tilde{V}_{i, t}) \rightarrow (\sum_{l=t}^{t+k-1} \gamma^{l-t} \tilde{r}_{i, l} + \gamma^{k} \tilde{V}_{i, t+k} - \tilde{V}_{i, t}$).

\paragraph{Mediator}\label{sec:algorithm_practical_mediator}

It is well-known that learning the joint policy in a centralized way is unfeasible due to the exponential scaling of the action space with the number of agents. For this reason, we fully factorize the joint policy of the mediator for a coalition: $\boldsymbol\pi^M_C(o_C^M) = \prod_{i \in C} \pi_i^M (o_i^M)$. The mediator's policies for all agents are parameterized with a single neural network $\theta^M$ that receives as input the observation $o_i^M = (o_i, C)$ and the agent's index $i$. For convenience, we denote the $i$-th agent policy $\pi_{\theta^M} ((o_i, C, i))$ as $\pi_{\theta_i^M} (o_i, C)$. The objectives for $\pi_{\theta_i^M}$ are discussed in the next subsection.

The mediator's critic receives coalition $C$ and observations of all agents $(o_i)_{i \in N}$ as an input and simultaneously outputs values for all agents, both in and out of the coalition. While this formulation requires centralized access to all observations and rewards, note that the critic is only required during training. Access to individual value functions for each agent will be crucial in the next subsection when estimating constrained objectives for the mediator. For convenience, we denote a critic's output $\tilde{V}_{\phi^M}(((o_i)_{i \in N}, C, i))$ as $\tilde{V}_{\phi_i^M}(\textbf{o}, C)$.  As usual, it is trained to minimize squared temporal difference:

\begin{equation}
\label{loss_mediator_critic}
L(\phi_i^M) = \mathbb{E} (\tilde{r}_{i, t} + \gamma \tilde{V}_{\phi_i^M}(\textbf{o}_{t+1}, C_{t+1}) - \tilde{V}_{\phi_i^M}(\textbf{o}_t, C_t))^2
\end{equation}

Note that our implementation is intentionally minimalistic. The literature suggests a multitude of sophisticated solutions for various problems typical for MARL. Instead of limiting the mediator to a fully factorized policy, sampling from a joint policy is possible with techniques based on coordination graphs \cite{guestrin2002multiagent,bohmer2020deep}. Credit assignment could be addressed by using specialized critics that isolate contributions of each agent to social welfare \cite{lowe2017multi,foerster2018counterfactual,ivanov2021balancing}. While these techniques could potentially improve the mediator, our focus is on introducing mediators to MARL (and in turn promoting the idea of applying mechanism design in MARL), not on searching for the best possible implementation of the mediator. We intend to isolate the implementation choices that are necessary for the mediator to function, and complicating the mediator is in conflict with this intent.

\subsection{Objectives for the Mediator}\label{sec:algorithm_objective}

Here, we first propose a mediator we denote as Naive that simply maximizes the social welfare of the agents in the coalition. We then derive the constraints necessary to incentivize the agents to commit, as well as the training procedure that approximately satisfies these constraints for $k=1$. Finally, we discuss how the training procedure should be modified for $k > 1$, as well as the potential effect of $k$.

\paragraph{Naive Mediators}\label{sec:algorithm_objective_naive}

The Naive mediator is oblivious to the incentives of the agents. Its goal is to greedily maximize the utilitarian social welfare for any given coalition. For the mediator's policy $\pi_i^M$ for some agent $i$, this goal can be formulated as the following objective:

\begin{equation}
\label{objective_naive}
\forall \textbf{o}_t, (C_t \mid i \in C_t): \max_{\pi_i^M} \sum_{j \in C_t} V_j(o_{j,t}, C_t)
\end{equation}

Each value function $V_j(o_{j,t}, C_t)$ is approximated with the $j$-th output of the mediator's critic $\tilde{V}_{\phi_j^M}(\textbf{o}_t, C_t)$. The policy $\pi_i^M$ of the mediator for the agent $i$ is trained via policy gradient:

\begin{equation}
\label{mediator_pg_naive}
\begin{split}
L(\theta_i^M) = -\mathbb{E} [  \sum_{j \in C_t} (\tilde{r}_{j, t} + \gamma \tilde{V}_{\phi_j^M} (\textbf{o}_{t+1}, C_{t+1}) & \\
- \tilde{V}_{\phi_j^M} (\textbf{o}_t, C_t) ) ] \log \pi_{\theta_i^M} (o_{i, t}, C_t) &
\end{split}
\end{equation}

\paragraph{Constraints}\label{sec:algorithm_objective_constraints}

Since the Naive mediator greedily optimizes social welfare, there is no guarantee that its optimal policy is a mediated equilibrium, i.e., that the agents always prefer committing. To fix this, the mediator's policy should satisfy certain constraints. Intuitively, self-interested agents only commit if committing serves their best selfish interests. In RL, the quantification of an agent's interests is a value function, so committing should yield a higher value than not committing for each agent. Below we mathematically express this condition. For convenience, we divide the agents into two groups: those in and those outside the coalition.

On the one hand, an agent that enters the coalition should benefit from it and receive payoffs at least as high as it (counterfactually) would outside the coalition:

\begin{equation}
\label{constraint_ic}
\forall i, o_{i, t}: \hspace{10pt} \mathbb{E}_{(C_t \mid i \in C_t)} [V_i(o_{i, t} \mid C_t)] \geq \mathbb{E}_{(C_t \mid i \in C_t)} [V_{i}(o_{i, t} \mid  C_t \setminus \{i\})]
\end{equation}

On the other hand, an agent that chooses to act on its own should not be able to exploit the mediator and should receive payoffs not higher than if it had (counterfactually) committed:

\begin{equation}
\label{constraint_p}
\forall i, o_{i, t}: \hspace{10pt} \mathbb{E}_{(C_t \mid i \notin C_t)} [V_i(o_{i, t} \mid  C_t \cup \{i\})] \geq \mathbb{E}_{(C_t \mid i \notin C_t)} [V_{i}(o_{i, t} \mid  C_t)]
\end{equation}

We respectively refer to the constraints (\ref{constraint_ic}) and (\ref{constraint_p}) as Incentive-Compatibility (IC) and Encouragement (E) constraints. Notice that they are in expectation over the distribution of coalitions (generated by the agents' policies), rather than for each coalition. This is because the agents choose actions before the coalition is formed.

Also, notice that the constraints are feasible since they can be exactly satisfied by the mediator copying the policies of the agents (thus having no effect). In other words, mediated equilibrium always exists. We now discuss how to train a mediator that satisfies these constraints while maximizing social welfare. 

\paragraph{Incentive-Compatibility Constraint}

To incorporate the constraints (\ref{constraint_ic}) into the objective (\ref{objective_naive}), we can apply the method of Lagrange multipliers, which results in the following dual objective:

\begin{equation}
\label{objective_ic_dual}
\forall \textbf{o}_t, (C_t \mid i \in C_t): \max_{\pi_i^M} \left[  \sum_{j \in C_t} V_j(o_{j,t}, C_t) + \lambda^{ic}_i(o_{i, t}) V_i(o_{i,t}, C_t) \right]
\end{equation}

\noindent where $\lambda^{ic}_i(o_{i, t}) \geq 0$ are Lagrange multipliers. Note that the counterfactual value $V_{i}(o_{i, t} \mid C_t \setminus \{i\})$ from (\ref{constraint_ic}) can be omitted from the dual due to not depending on $\pi_i^M$. This objective can be maximized using policy gradient:

\begin{equation}
\label{mediator_pg_ic}
\begin{split}
L(\theta_i^M) = - \mathbb{E} [ \sum_{j \in C_t} (\tilde{r}_{j, t} + \gamma \tilde{V}_{\phi_j^M} (\textbf{o}_{t+1}, C_{t+1}) -  \tilde{V}_{\phi_j^M} (\textbf{o}_t, C_t)) & \\
+ \lambda^{ic}_i (o_{i, t}) (\tilde{r}_{i, t} + \gamma \tilde{V}_{\phi_i^M} (\textbf{o}_{t+1}, C_{t+1}) - \tilde{V}_{\phi_i^M} (\textbf{o}_t, C_t)) &] \\
\log \pi_{\theta_i^M} (o_{i, t}, C_t) &
\end{split}
\end{equation}

To mitigate credit assignment, we deliberately incorporate only the IC constraint of the $i$-th agent into the objective of $\pi_i^M$ but not the IC constraints of all agents. This way, the mediator only has to ensure that its actions for an agent are compatible with the incentives of this agent.

To find $\lambda^{ic}_i(o_{i, t})$, we employ dual gradient descent \cite{boyd2004convex}:

\begin{equation}
    \label{lambda_ic}
    \log \lambda^{ic}_i(o_{i, t}) \leftarrow \log \lambda^{ic}_i(o_{i, t}) - \alpha [\tilde{V}_{\phi_i^M}(\textbf{o}_t, C_t) - \tilde{V}_{\phi_i^M}(\textbf{o}_t, C_t \setminus \{i\})]
\end{equation}

\noindent where $\alpha$ is the learning rate. The intuition behind this update is that the second term on the right-hand side on average equals zero when the constraint is satisfied exactly. In practice, we update $\lambda^{ic}_i$ once according to this update each time we update actors and critics. To ensure that $\lambda^{ic}_i$ is non-negative, we update its logarithm instead. Furthermore, in our experiments, we find approximation as a scalar $\lambda^{ic}_i(o_{i, t}) = \lambda^{ic}_i$ to be sufficient. Other examples of applying dual gradient descent to enforce a constraint can be found in \cite{peng2018variational,ivanov2022optimal}. 

Note that the update rule (\ref{lambda_ic}) involves a counterfactual value $\tilde{V}_{\phi_i^M}(\textbf{o}_t, C_t \setminus \{i\})$, i.e., what the value of the $i$-th agent would be if it did not enter this specific coalition in this specific state. Estimating it is only possible thanks to our implementation of the mediator's critic, i.e., our choice to train it to simultaneously estimate values of agents both in and out of the coalition (see Section \ref{sec:algorithm_practical_mediator}).

\paragraph{Encouragement Constraint} Same derivations as for the IC constraint apply to the E constraint. The constraints (\ref{constraint_p}) for each agent outside the coalition can be incorporated into the policy gradient (\ref{mediator_pg_naive}) of each agent in the coalition $i \in C_t$:

\begin{equation}
\label{mediator_pg_p}
\begin{split}
L(\theta_i^M) = - \mathbb{E} [  \sum_{j \in C_t} (\tilde{r}_{j, t} + \gamma \tilde{V}_{\phi_j^M} (\textbf{o}_{t+1}, C_{t+1}) - \tilde{V}_{\phi_j^M} (\textbf{o}_t, C_t) ) & \\
- \sum_{j \notin C_t} \lambda^{e}_j (o_{j, t}) (\tilde{r}_{j, t} + \gamma \tilde{V}_{\phi_j^M} (\textbf{o}_{t+1}, C_{t+1}) - \tilde{V}_{\phi_j^M} (\textbf{o}_t, C_t)) & ] \\
\log \pi_{\theta_i^M} (o_{i, t}, C_t) &
\end{split}
\end{equation}

The Lagrange multipliers $\lambda^{e}_j(o_{i, t})$ are also learned via dual gradient descent:

\begin{equation}
    \label{lambda_p}
    \log \lambda^{e}_j(o_{j, t}) \leftarrow \log \lambda^{e}_j(o_{j, t}) - \alpha [\tilde{V}_{\phi_j^M}(\textbf{o}_t, C_t \cup \{j\}) - \tilde{V}_{\phi_j^M}(\textbf{o}_t, C_t)]
\end{equation}

The Lagrange multipliers are also as scalars: $\lambda^{e}_j(o_{i, t})=\lambda^{e}_j$.

\paragraph{Constrained mediators}\label{sec:algorithm_objective_both}

Both IC and E constraints should be applied simultaneously to train the Constrained mediator.

\begin{equation}
\label{mediator_pg_both}
\begin{split}
L(\theta_i^M) = - \mathbb{E} [ \sum_{j \in C_t} (\tilde{r}_{j, t} + \gamma \tilde{V}_{\phi_j^M} (\textbf{o}_{t+1}, C_{t+1}) -  \tilde{V}_{\phi_j^M} (\textbf{o}_t, C_t)) & \\
+ \lambda^{ic}_i (o_{i, t}) (\tilde{r}_{i, t} + \gamma \tilde{V}_{\phi_i^M} (\textbf{o}_{t+1}, C_{t+1}) - \tilde{V}_{\phi_i^M} (\textbf{o}_t, C_t)) & \\
- \sum_{j \notin C_t} \lambda^e_j (o_{j, t}) (\tilde{r}_{j, t} + \gamma \tilde{V}_{\phi_j^M} (\textbf{o}_{t+1}, C_{t+1}) - \tilde{V}_{\phi_j^M} (\textbf{o}_t, C_t)) & ] \\
\log \pi_{\theta_i^M} (o_{i, t}, C_t) &
\end{split}
\end{equation}

It is interesting to note that the loss (\ref{mediator_pg_both}), obtained as a dual of a constrained objective, is also a (negated) policy gradient for a mixture of rewards $[\sum_{j \in C_t} \tilde{r}_{j, t} + \lambda^{ic}_{i, t} \tilde{r}_{i, t} - \sum_{j \notin C_t} \lambda^e_{j, t} \tilde{r}_{j, t}]$.\footnote{To see this, apply the definition of the value function, change the order of expectation and summation in the value functions, and rearrange terms.} One implication is that socially beneficial equilibria can be found by simply optimizing a weighted sum of rewards, albeit with non-stationary weights. Another implication is that the E constraint effectively lowers the rewards of agents outside the coalition. We stress that this does not mean that the agents outside the coalition are punished, but rather that the agents inside the coalition cooperate less frequently if the coalition is not full. This ensures that an agent cannot deviate to enjoy the cooperation of others, i.e., deters free-riding.

\paragraph{Commitment Window $k$}

When $k=1$, unanimous commitment requires both constraints to be satisfied at each time-step, which may limit the margin of social welfare improvement over selfish agents. In contrast, when $k > 1$, the constraints only need to be satisfied on average over the periods of $k$ time-steps, since agents commit for these periods. On the example of IC constraint (\ref{constraint_ic}), $k > 1$ requires constraint reformulation:

\begin{equation}
    \label{constraint_ic_k}
    \begin{split}
    & \forall i, (o_{i, t} \mid t \mod k = 0): \\
    & \hspace{10pt} \mathbb{E}_{(C_t \mid i \in C_t)} [V_i(o_{i, t} \mid C_t)] \geq \mathbb{E}_{(C_t \mid i \in C_t)} [V_{i}(o_{i, t} \mid C_t \setminus i)]
    \end{split}
\end{equation}

\noindent where the coalition $C_t$ is fixed for the next $k$ time-steps: $C_t = C_{t+1} = \dots = C_{t+k-1}$. This constraint implies that the same $\lambda^{ic}_i(o_{i, t})$ is used in the dual objective (\ref{mediator_pg_ic}) at the time-steps $t \leq l < t + k$. The update rules of the Lagrange multipliers are modified accordingly:

\begin{equation}
    \label{lambda_ic_k}
    \begin{split}
    \log & \lambda^{ic}_i(o_{i, t}) \leftarrow \log \lambda^{ic}_i(o_{i, t}) \\ 
    & - \alpha \sum_{l=t}^{t+k-1} \gamma^{l-t} [\tilde{V}_{\phi_i^M}(\textbf{o}_l, C_l) - \tilde{V}_{\phi_i^M}(\textbf{o}_l, C_l \setminus \{i\})]
    \end{split}
\end{equation}

\begin{equation}
    \begin{split}
    \label{lambda_p_k}
    \log & \lambda^{e}_j(o_{j, t}) \leftarrow \log \lambda^{e}_j(o_{j, t}) \\
    & - \alpha \sum_{l=t}^{t+k-1} \gamma^{l-t} [\tilde{V}_{\phi_j^M}(\textbf{o}_t, C_t \cup \{j\}) - \tilde{V}_{\phi_j^M}(\textbf{o}_t, C_t)]
    \end{split}
\end{equation}

In the extreme case when the commitment window covers the entire episode ($k = \inf$) and the starting state $s_0$ is deterministic, the dependency of Lagrange multipliers on the observation can be dropped altogether. We respectively denote the mediators that require constraint satisfaction each time-step ($k=1$), each several time-steps ($k>1$), and each episode ($k=\inf$) as \textit{ex-post}, \textit{interim}, and \textit{ex-ante}.

\paragraph{Notes on the training process.}

First, in our implementations of the mediator, no specific learning dynamics are assumed to be adopted by the agents. We only train the agents within the same RL framework as the mediator out of convenience. The only assumption is that the agents are myopically rational, i.e., increase the probability of the beneficial actions as they learn (and thus commit when are properly incentivized to do so). This is unlike approaches to conditional cooperation like LOLA that assume specific learning dynamics.

Second, because the mediator's policy is trained \textit{for each coalition} rather than only the full coalition (see equations \ref{objective_naive}, \ref{objective_ic_dual}), our objective formulation for the mediator is stronger than finding a socially beneficial mediated equilibrium. Instead, the mediator learns to maximize social welfare while satisfying constraints even for coalitions that are not full. On the one hand, this makes cooperation induced by the mediator robust to rare defectors. On the other hand, discovering that simultaneous commitment is beneficial requires some (possibly spontaneous) coordination of the agents, which is easier if non-unanimous commitment is also incentivized. In short, this property is useful during both training and execution. 

\section{Experiments}

In this section, we experimentally investigate the capabilities of different proposed versions of the mediator in Prisoner's Dilemma (PD), Public Good Game (PGG), and iterative PGG. Technical details and hyperparameters, as well as additional experiments, are reported in the Appendix.

In the Introduction, we have extensively discussed the issues that arise in mixed MARL, i.e., finding policies that maximize social welfare and incentivizing agents to follow these policies, which we respectively denote as Efficiency (Eff) and Incentive-Compatibility (IC) for convenience. While unconditional cooperation only addresses Eff, conditional cooperation addresses both and thus is substantially more difficult to resolve. Given that unconditional cooperation is already widely explored in the literature, we primarily focus on IC. To this end, we empirically investigate matrix games and their sequential variants that are trivial from the perspective of Eff but clearly highlight the specific challenges when dealing with IC, as well as the power of mediators in dealing with these challenges.

A consequence of this decision is that comparing with baselines becomes redundant. We design each investigated game such that we know what the socially optimal policy looks like and how much social welfare it achieves. By design, this policy is trivial to find for any solution to unconditional cooperation. We note that the next logical step is validating our mediators in more complex environments where both Eff and IC issues are non-trivial to solve, in which case comparing with baselines also makes sense. We leave this direction as future work.

\subsection{Prisoner's Dilemma}

The payoff matrix is presented in Table \ref{table:pd} (a). Despite the cooperation being mutually beneficial, defecting is the dominant strategy and mutual defection is the only equilibrium. Table \ref{table:pd} (b) presents socially optimal mediated equilibrium in PD implemented by a mediator that cooperates only when both agents commit. Notice that despite this mediator greedily maximizing social welfare for each coalition, both IC and E constraints are satisfied for both agents because committing is a dominant strategy. Therefore, one can expect even a Naive mediator to establish cooperation.

The results are presented in Table \ref{table:pd_results}. In accordance with our expectations, in the absence of the mediator, both agents converge to defection. When the game is augmented with a Naive mediator, the agents almost exclusively commit, and the mediator learns to only cooperate when both agents commit. 

\begin{table}[t]
\caption{Results in one-step PD. $c$ denotes Cooperate in PD and Contribute in PGG; $m$ denotes Commit; $\pi$ and $\pi^M$ respectively denote policies of agents and mediator; $\tilde{\pi}$ denotes policy averaged over 100 sampled episodes; reward is normalized between 0 and 1. The same caption applies to other tables. \\ \centerline{$^{(*)}$ no mediator $^{(\dagger)}$ naive mediator}}
\label{table:pd_results}
\centering
\begin{tabular}[t]{@{}lcc@{}}
\toprule
         \textbf{PD}                  & agent 0 & \multicolumn{1}{l}{agent 1} \\ \midrule
 $\pi(c)^{(*)}$           & 0.004  & 0.001                      \\
$\pi(m)^{(\dagger)}$           & 0.96    & 0.967                       \\
$\pi(c)^{(\dagger)}$           & 0.006  & 0.003                      \\
$\pi^M(c \mid [0, 1])^{(\dagger)}$ & -     & 0.01                      \\
$\pi^M(c \mid [1, 0])^{(\dagger)}$  & 0.009  & -                         \\
$\pi^M(c \mid [1, 1])^{(\dagger)}$  & 0.979  & 0.979                       \\ \bottomrule
\end{tabular}
\end{table}



\subsection{Public Good Game}

In two-player games, the mediator either acts for one of the agents, in which case the best it can do is maximize the agent's welfare, or acts for both agents, in which case no agent is outside the coalition. The consequence is that mediator's actions are never beneficial for agents outside the coalition and therefore the Encouragement constraint is always satisfied in two-agent games. This is not the case in games with more than two agents where as soon as some two agents start cooperating, the rest of the agents may try to exploit their cooperation for higher personal gains. This issue is known as free-riding. Using PGG as an example, we illustrate how free-riding emerges when the game is augmented with a Naive mediator, and how this issue can be mitigated by enforcing the Encouragement constraint on the mediator's policy. We refer to such a mediator as Constrained. We first investigate one-step PGG and then move on to our variant of iterative PGG.

\begin{table}[t]
\captionsetup{justification=centering}
\caption{Results in one-step PGG \\ $^{(*)}$ no mediator $^{(\dagger)}$ naive mediator $^{(\ddagger)}$ constrained mediator}
\label{table:pgg_results}
\centering
\begin{tabular}[t]{@{}llccc@{}}
\toprule
\textbf{PGG}            & \multicolumn{1}{l}{$N=3$}  & \multicolumn{1}{l}{$N=10$} & \multicolumn{1}{l}{$N=25$}\\ \midrule
reward$^{(*)}$  &  0.012  &  0.0  &   0.0  \\
reward$^{(\dagger)}$  &  0.652  &  0.005  &   0.014 \\
$\tilde{\pi}(m)^{(\dagger)}$      &   0.658  &  0.159  &  0.121  \\
$\tilde{\pi}^{M}(c)^{(\dagger)}$  &   0.985  &  0.001  &  0.02  \\
$\pi^M(c \mid |C| = 2)^{(\dagger)}$  & 0.993 & - & - \\ 
$\pi^M(c \mid |C| = 3)^{(\dagger)}$  & 0.999 & - & - \\
reward$^{(\ddagger)}$ & 0.891 & 0.827  & 0.817 \\
$\tilde{\pi}(m)^{(\ddagger)}$     & 0.916 & 0.961 & 0.933 \\
$\tilde{\pi}^{M}(c)^{(\ddagger)}$ & 0.959 & 0.858 & 0.817 \\
$\pi^M(c \mid |C| = 2)^{(\ddagger)}$  & 0.774 & - & - \\ 
$\pi^M(c \mid |C| = 3)^{(\ddagger)}$  & 0.996 & - & - \\
\bottomrule
\end{tabular}
\end{table}

\paragraph{One-step Public Good Game}

$N$ agents are endowed with a unit of utility and have a choice whether to contribute it to the public good or to defect. The public good is formed as the total contribution of agents, multiplied by some $1 < n < N$, and is uniformly redistributed among all agents. The reward of each agent is $r_i = \frac{n}{N} \sum_{j=1}^N c_j - c_i$, where $c_i=1$ iff $i$ contributes.

Let $N=3, n=2$. Like in PD, the dominant strategy in the absence of a mediator is to defect. Consider the Naive mediator. Its strategy is to contribute with all agents in the coalition if it consists of at least two agents: $\pi^M(c \mid |C| = 1)=0, \pi^M(c \mid |C| \geq 2)=1$. Given this mediator, the equilibrium is for only two of three agents to commit and form a coalition. To see this, consider the perspectives of all agents. From the perspective of an agent in this coalition, committing causes it to get its share of the public good equal to $r(m)=1/3$, whereas defecting would lower the reward to $r(d)=0$ by causing the other agent to defect. From the perspective of the agent outside the coalition, defecting lets it enjoy both its endowment and its share of the public good $r(d)=4/3$, which is better than committing with other agents and receiving $r(m)=1$. Now consider the Constrained mediator. To deter free-riding, the Constrained mediator contributes with a specific probability when the coalition consists of two agents: $\pi^M(c \mid |C| = 1)=0, \pi^M(c \mid |C| = 2)=0.75, \pi^M(c \mid |C| = 3)=1$. On the one hand, this results in a lower expected reward for the two agents in the coalition: $r(m)=0.25$. On the other hand, this also lowers the expected reward of the agent outside the coalition to the point where it is indifferent whether it commits or not: $r(d) = r(m) = 1$. Notice that further decreasing $\pi^M(c \mid |C| = 2)$ results in lower social welfare for the agents in the coalition, whereas increasing it encourages the third agent to defect. Thus, this mediator implements the socially optimal equilibrium. We now verify that our constrained objective allows training such a mediator.

The results are presented in Table \ref{table:pgg_results}. For $N=3, n=2$, the learned policies match the equilibrium derived above: without mediator agents always defect, the Naive mediator encourages two agents to commit but is exploited by the third agent, and the Constrained mediator converges to the socially optimal equilibrium. It is especially surprising that the Constrained mediator learns the optimal mixed policy so precisely, which is only possible in a non-stationary environment where the moment the mediator deviates, it is corrected by the agents trying to exploit it. For settings $N=10, n=2$ and $N=25, n=5$, the picture is generally the same: only the Constrained mediator encourages commitment from all agents by learning a reciprocal policy that punishes free-riding. 

\begin{table}[t]
\captionsetup{justification=centering}
\caption{Results in Iterative PGG. The reported policies are empirical approximations of the marginal probabilities. \\ $^{(*)}$ no mediator $^{(\dagger)}$ naive mediator $^{(\mathcal{\ddagger})}$  constrained mediator}
\label{table:pgg_iter_results}
\centering
\begin{tabular}{ cc }
\begin{tabular}{@{}llcc@{}}
\toprule
\textbf{N=3}            & $k=1$  & $k=10$ \\ \midrule                              
reward$^{(*)}$               & 0.019                     & 0.017                     \\
reward$^{(\dagger)}$               & 0.145                     & 0.963                     \\
    $\tilde{\pi}(m)^{(\dagger)}$    & 0.667 & 0.991                     \\
    $\tilde{\pi}^{M}(c)^{(\dagger)}$ & 0.999 & 0.993                     \\
reward$^{(\mathcal{\ddagger})}$ & 0.478 & 0.986 \\
                                         $\tilde{\pi}(m)^{(\mathcal{\ddagger})}$     & 0.8 & 0.995 \\
                                         $\tilde{\pi}^{M}(c)^{(\mathcal{\ddagger})}$ & 0.997 & 0.999 \\ \bottomrule
\end{tabular} & 
\begin{tabular}{@{}llcc@{}}
\toprule
\textbf{N=10}            & $k=1$  & $k=10$ \\ \midrule
reward$^{(*)}$               & 0.0                    & 0.0                     \\
reward$^{(\dagger)}$                & 0.0                     & 0.788                     \\
                                        $\tilde{\pi}(m)^{(\dagger)}$    & 0.079                     & 0.852                     \\
                                        $\tilde{\pi}^{M}(c)^{(\dagger)}$ & 0.159                     & 0.932                     \\
reward$^{(\mathcal{\ddagger})}$ & 0.729 & 0.787 \\
                                        $\tilde{\pi}(m)^{(\mathcal{\ddagger})}$     & 0.898 & 0.907 \\
                                        $\tilde{\pi}^{M}(c)^{(\mathcal{\ddagger})}$ & 0.945 & 0.941 \\ \bottomrule
\end{tabular}
\end{tabular}
\end{table}

\paragraph{Iterative Public Good Game}

The game lasts for 10 turns. In the beginning, each agent is endowed with 1 unit of utility. An agent's observation is a tuple of its current endowment and the turn number. Each turn, each agent chooses whether to contribute $50\%$ of its current endowment to the public good, and the resulting payoffs are preserved throughout the turns. This creates a compounding effect from contributing to the public good that can be exploited for a massive increase in welfare over the duration of the game if all agents consistently contribute. On the other hand, the state space is no longer trivial, and more complex strategies may emerge.

The results are presented in Table \ref{table:pgg_iter_results}. For $N=3, n=2$, the Naive ex-post ($k=1$) mediator behaves similarly to the Naive mediator in one-step PGG: it consistently encourages two agents to commit but is exploited by the third agent. The Constrained ex-post mediator mitigates this issue, but only partially, which might be due to our approximation of Lagrange multipliers as constants, or simply due to the limited capabilities of ex-post mediators. Conversely, both the Naive and the Constrained ex-ante ($k=10$) mediators reliably encourage all three agents to commit and establish robust cooperation. For $N=10, n=5$, the results are similar, but since the game is more complex, even the Constrained ex-ante mediator is not able to ensure full commitment.

\subsection{Prisoner's Dilemma with Sacrifice}

Prisoner's Dilemma with Sacrifice (PDS) is an asymmetric modification of PD. The payoff matrix is presented in Table \ref{table:pd_sac} and differs from PD in that the second player has an additional action available, the effect of which is to sacrifice its payoffs for the higher utilitarian social welfare.

Like in PD, when agents play PDS without a mediator, the dominant action for both is to defect. Unlike PD, this does not change when the game is augmented with the Naive mediator. Since the Naive mediator greedily maximizes social welfare, it sacrifices the payoffs of the second agent, which encourages the second agent to defect (Table \ref{table:pd_sac_naive_med}). This is an example of the incompatibility of an agent's and the mediator's incentives. To fix this, the IC constraint should be enforced, which will cause the mediator to choose mutual cooperation over sacrificing an agent's payoffs and, in turn, encourage the agents to commit (Table \ref{table:pd_sac_ic_med}). This Constrained mediator implements the mediated equilibrium, but due to the asymmetry of the game, its strategy is not socially optimal. Note that this is not the only mediator that satisfies the IC constraint, as mixing mutual cooperation with sacrificing the second agent's payoffs may also be viable while further improving social welfare. The mediated equilibrium that maximizes social welfare is to equally mix (c, c) and ($\cdot$, s) outcomes since at this point the second agent is indifferent to whether it commits or defects.

\begin{table}[t]
\caption{Payoffs in PD with Sacrifice}
\label{table:pd_sac}
\centering
\begin{tabular}{@{}lccc@{}}
\toprule
          & Defect & Cooperate & Sacrifice \\
Defect    & 1, 1   & 3, 0     & 5, 0     \\
Cooperate & 0, 3  & 2, 2      & 5, 0     \\ \bottomrule
\end{tabular}
\end{table}

\begin{table}[t]
\captionsetup{justification=centering}
\caption{Payoffs in PD with Sacrifice augmented with a Naive mediator, the strategy of which is to sacrifice the second agent's payoffs when both agents commit and defect otherwise}
\label{table:pd_sac_naive_med}
\centering
\begin{tabular}{@{}lcccc@{}}
\toprule
          & Defect & Cooperate & Sacrifice & Commit \\
Defect    & 1, 1   & 3, 0     & 5, 0  & 1, 1     \\
Cooperate & 0, 3   & 2, 2     & 5, 0  & 0, 3    \\
Commit    & 1, 1   & 3, 0     & 5, 0  & 5, 0    \\
\bottomrule
\end{tabular}
\end{table}

\begin{table}[t]
\captionsetup{justification=centering}
\caption{Payoffs in PD with Sacrifice augmented with a Constrained mediator, the strategy of which is to mutually cooperate when both agents commit and defect otherwise}
\label{table:pd_sac_ic_med}
\centering
\begin{tabular}{@{}lcccc@{}}
\toprule
          & Defect & Cooperate & Sacrifice & Commit \\
Defect    & 1, 1   & 3, 0     & 5, 0  & 1, 1     \\
Cooperate & 0, 3   & 2, 2     & 5, 0  & 0, 3    \\
Commit    & 1, 1   & 3, 0     & 5, 0  & 2, 2    \\
\bottomrule
\end{tabular}
\end{table}

We now investigate how our implementations of mediators behave in PDS. The experimental results are presented in Table \ref{table:pd-sac-res}.

In accordance with our expectations, agents converge to mutual defection both without a mediator and with a Naive mediator. The Naive mediator learns to sacrifice the second agent's payoffs while defecting with the first agent, which causes the second agent to always defect. The first agent is then indifferent to whether it defects itself or commits to the mediator that defects for it.

The constrained mediator performs much better. As discussed earlier, its optimal strategy is to equally mix (c, c) and ($\cdot$, s) outcomes. Its learned strategy is close to the optimal but gives a slight edge to the (c, c) outcome to additionally encourage the commitment of the second agent. As a result, both agents almost always commit. The converged dynamics result in social welfare of approximately 4.35, which is close to the maximal achievable social welfare of 4.5.

\begin{table}[t]
\captionsetup{justification=centering}
\caption{Learned policies in PDS. Note that the mediator's joint policy is not factorized. $\{\cdot\}$ denotes any action. \\ $^{(*)}$ no mediator $^{(\dagger)}$ naive mediator $^{(\ddagger)}$ constrained mediator}
\label{table:pd-sac-res}
\centering
\begin{tabular}{cccc}
\toprule
 & & \multicolumn{1}{l}{agent 0} & \multicolumn{1}{l}{agent 1} \\
 \midrule
agents$^{(*)}$ & $\pi(c)$  & 0.002  & 0.001 \\ 
agents$^{(\dagger)}$   & $\pi(c)$  & 0.003 & 0.002 \\
                                        & $\pi(m)$  & 0.607 & 0.002 \\
mediator$^{(\dagger)}$ & $\pi^M(c, c | {[}1,1{]})$ & \multicolumn{2}{c}{0.001} \\
                       & $\pi^M(\cdot, s | {[}1,1{]})$ & \multicolumn{2}{c}{0.85}\\
agents$^{(\ddagger)}$  & $\pi(c)$ & 0.001 & 0.001 \\
                                        & $\pi(m)$ & 0.995 & 0.982 \\
mediator$^{(\ddagger)}$ & $\pi^M(c, c | {[}1,1{]})$ & \multicolumn{2}{c}{0.601} \\
                       & $\pi^M(\cdot, s | {[}1,1{]})$ & \multicolumn{2}{c}{0.398}\\
\bottomrule
\end{tabular}
\end{table}

\section{Conclusion}

In this paper, we challenge the dominant perspective in the MARL literature on the problem of cooperation in mixed environments and argue for convergence to equilibria as its essential property. As a novel solution for conditional cooperation, we apply mediators. Specifically, we adapt mediators to Markov games through the formalism of Markov mediators, describe how to practically implement them, formulate a constrained objective that both improves social welfare and encourages agents to commit, solve this objective using the method of Lagrange multipliers and dual gradient descent, and experimentally verify the effectiveness of our implementation in the matrix and iterative games.

Despite our contributions, we only scratch the surface of the mediators' potential for MARL. First, to get a clear picture of mediators' behaviour and advantages, we experiment with relatively simple games, but it would also be exciting to apply mediators to larger-scale environments. Second, our formulation of mediator implies its centralized execution as a way to ensure that agents cannot misreport their commitment, but as \citet{monderer2009strong} point out, it is interesting whether cryptographic technologies could be applied as an alternative. Third, in our implementation, the mediator acts based on the same information as an agent, but providing the mediator with more information could serve as an additional incentive to commit. Fourth, the literature also explores mediators that recommend actions instead of acting on behalf of agents \cite{kearns2014mechanism} and adapting such mediators to MARL presents a separate challenge. On a final note, our Markov mediator is only one example of ideas from economics to MARL, and we are excited for future research that intersects these two fields.  

\begin{acks}
This research was supported in part through computational resources of HPC facilities at HSE University, Russian Federation. Support from the Basic Research Program of the National Research University Higher School of Economics is gratefully acknowledged.
\end{acks}

\balance
\bibliographystyle{aamas2023/ACM-Reference-Format} 
\bibliography{main}

\appendix

\section{Additional Experiments}

\subsection{Two-step Asymmetric Prisoner's Dilemma}

This modification of PD lasts for two time-steps, the payoff matrix for both of which is provided in Table \ref{table:two-step-pd}. The second state coincides with PD in the main text, but the first state is different in that mutual cooperation is only beneficial for the second agent while still providing maximal social welfare. Like in one-step PD, the only equilibrium is to defect for both agents in the absence of mediator. The ex-post Naive mediator chooses mutual cooperation in both states, which both agents agree to in the second state but only the second agent agrees to in the first state. The ex-ante Naive mediator also chooses mutual cooperation in both states, but the agents can only commit at the first state for the duration of the game. In this case, commitment is beneficial for both agents since the cumulative reward over two time-steps is higher from mutual cooperation than from mutual defection even for the first agent.

The experimental results are presented in Table \ref{table:two-step-pd-res} and are fully in accordance with our expectations. Agents always defect without mediator; the second agent commits in both states while the first agent only commits in the second state to the ex-post Naive mediator; both agents commit in the first state to the ex-ante Naive mediator. This experiment clearly demonstrates how ex-ante mediator has more potential to maximize social welfare because it only requires to satisfy the constraints (to be compatible with the agents' incentives) on average.

\begin{table}[t]
\caption{Two-step Prisoner's Dilemma}
\label{table:two-step-pd}
\centering
\begin{tabular} { cc }
\begin{tabular}{@{}lcc@{}}
\toprule
STATE 0   & Defect & Cooperate \\
Defect    & 0, 0   & 7, -5     \\
Cooperate & -5, 7  & \textbf{-1, 4}     \\ \bottomrule
\end{tabular} &
\begin{tabular}{@{}lcc@{}}
\toprule
STATE 1   & Defect & Cooperate \\
Defect    & 0, 0   & 7, -5     \\
Cooperate & -5, 7  & 2, 2      \\ \bottomrule
\end{tabular}
\end{tabular}
\end{table}

\begin{table}[ht]
\captionsetup{justification=centering}
\caption{Learned policies in Two-step PD\\ $^{(*)}$ no mediator $^{(\dagger)}$ ex-post naive mediator ($k=1$) $^{(\ddagger)}$ ex-ante naive mediator ($k=2$)}
\label{table:two-step-pd-res}
\centering
    \begin{tabular}{@{}llc@{}}
\toprule
\textbf{STATE 0}                                            & agent 0 & agent 1 \\ \midrule
$\pi(C)^{(*)}$              & 0.0007                      & 0.0006                      \\
$\pi(C)^{(\dagger)}$              & 0                           & 0                           \\
$\pi(M)^{(\dagger)}$             & \textbf{0.0128}                      & 0.9996                       \\
$\pi^M(C \mid {[}0,1{]})^{(\dagger)}$ & —                         & 0.0005                      \\
$\pi^{M}(C \mid {[}1,0{]})^{(\dagger)}$ & 0.0088                      & —                         \\
$\pi^{M}(C \mid {[}1,1{]})^{(\dagger)}$ & 0.9987                      & 0.9976                      \\
$\pi(C)^{(\ddagger)}$             & 0                           & 0                           \\
$\pi(M)^{(\ddagger)}$             & \textbf{0.9913}                      & 0.9995                      \\
$\pi^{M}(C \mid {[}0,1{]})^{(\ddagger)}$ & —                         & 0.0624                      \\
$\pi^{M}(C \mid {[}1,0{]})^{(\ddagger)}$ & 0.027                       & —                         \\
$\pi^{M}(C \mid {[}1,1{]})^{(\ddagger)}$ & 0.9985                      & 0.9982                       \\ \bottomrule
\end{tabular}

\begin{tabular}{@{}llc@{}}
\toprule
\textbf{STATE 1}                          & agent 0 & agent 1 \\ \midrule
$\pi(C)^{(*)}$              & 0.0005                      & 0.0006                      \\
$\pi(C)^{(\dagger)}$              & 0                           & 0.0001                      \\
$\pi(M)^{(\dagger)}$             & 0.9966                      & 0.9293                      \\
$\pi^{M}(C \mid {[}0,1{]})^{(\dagger)}$ & —                         & 0.0003                      \\
$\pi^{M}(C \mid {[}1,0{]})^{(\dagger)}$ & 0.0067                      & —                         \\
$\pi^{M}(C \mid {[}1,1{]})^{(\dagger)}$ & 0.9993                      & 0.9977                      \\
$\pi(C)^{(\ddagger)}$              & 0                           & 0                           \\
$\pi(M)^{(\ddagger)}$             & —                      & —                      \\
$\pi^{M}(C \mid {[}0,1{]})^{(\ddagger)}$ & —                         & 0.0028                      \\
$\pi^{M}(C \mid {[}1,0{]})^{(\ddagger)}$ & 0.0154                      & —                         \\
$\pi^{M}(C \mid {[}1,1{]})^{(\ddagger)}$ & 0.9948                      & 0.9933                      \\ \bottomrule
\end{tabular}
\end{table}

\section{Technical Details and Hyperparameters}


\paragraph{Prisoner's Dilemma}\label{para:pd}
In PD, agents' actor and critic receive a constant dummy state, since it is a one-step game with single state. Mediator's actor also receives coalition and ID of the agent that the mediator acts for. Mediator's critic only receives coalition and predicts values for both agents simultaneously. The algorithm of inference is the same in all environments: first, agents choose an action, then if any of them chose to cooperate, its ID alongside with coalition are passed to the mediator, which takes actions for these agents. After that, actions are sent to the environment to obtain rewards. The training is performed in the usual manner for Actor-Critic algorithms. The final result is averaged over 50 seeds.

\paragraph{Prisoner's Dilemma with Sacrifice}\label{para:sac}

We bound $\log \lambda$ to $[-4; 4]$ to avoid cases when the constraint is completely ignored or completely dominates the main objective. The rest of the details are similar to the \hyperref[para:pd]{PD}. The final result is averaged over 50 seeds.

\paragraph{Two-step Asymmetric Prisoner's Dilemma}\label{para:tspd}
Agents' actors and critics receive the time-step $t$. Also, actors receive an additional value $s \in [-1, 0, 1]$ that indicates the coalition status of an agent: ["cannot join the coalition", "can choose to join the coalition", "in coalition, acts according to mediator"]. Depending on the value, we modify the logits predicted by actor according to the following strategies. For $s=-1$, we mask the value corresponding to the action "commit" by changing it to $-\infty$ (this happens when $k>1$, $t \mod k \neq 0$, and the agent did not commit the last time it could). The same applies for $s=1$, but in this case we mask all actions but "commit" (this happens when $k>1$, $t \mod k \neq 0$, and the agent committed the last time it could). In case of $s=0$, neither mask is applied as all actions are available. During training, we mask logits in the same manner according to the collected trajectories to ensure unbiased on-policy learning. Agents are only trained on the experience where $s=-1$ or $s=0$ because when $s=1$, agents effectively act off-policy (as mediator acts for them). Note that $s = 0$ always if $k = 1$.

Since we use ex-ante mediator, we employ a $k$-step learning procedure for agents explained in the main text under section "Practical Implementations of Agents and Mediator". The rest of the details are similar to the \hyperref[para:pd]{PD}. The final result is averaged over 50 seeds.

\paragraph{Public Good Game}
Considering the high number of agents in PGG $N\in\{3, 10, 25\}$, we changed the multi-headed mediator's critic. Instead, the critic takes only the number of agents in coalition (normalized by $N$), and outputs value $V$ corresponding to each agent in the coalition. Likewise, the mediator's actor also doesn't return a unique policy for each agent. Instead, it outputs the same policy for all agents in the coalition. This way, we utilize the symmetry of the game to reduce the space of solutions. It is important to note that each agent still has its own actor and critic networks that do not share parameters with other agents. The rest of the details are similar to the \hyperref[para:pd]{PD}. The final result is averaged over 10 seeds.

\paragraph{Iterative Public Good Game}
In the Iterative PGG, we provide the private observation $o_{i, t} = (e_{i, t}, t)$ to agents' actors and critics, where $e_{i, t}$ is the $i$-th agent current endowment. Mediator's actor receives a tuple $(o_{i, t}, C_t, i)$ consisting of agent's private observation, coalition, and agent's ID, and outputs the policy for this agent. Mediator's critic receives a tuple $(s_t=(o_{i, t})_{i \in N}, C)$ consisting of the global state and the coalition and returns a vector of values $V$ of all agents. The rest of the details (including masking logits for $k > 1$) are the same as in \hyperref[para:tspd]{Two-step PD}. The final result is averaged over 10 seeds.

\paragraph{Hyperparameters}

All hyperparameters are reported in Table \ref{table:hp}.

\begin{table*}[t]
\caption{Hyperparameters for all environments}
\label{table:hp}
\centering
\begin{tabular}{@{}lccccc@{}}
\toprule
\multicolumn{1}{c}{}   & PD     & Two-step PD & Sacrifice PD & PGG         & PGG-iter    \\ \midrule
Batch Size             & 128    & 128         & 128          & 128         & 128         \\
Iterations             & 2000   & 2000        & 10000        & 20000       & 20000       \\ 
Discount factor $\gamma$                   & 0.99   & 0.99        & 0.99         & 0.99        & 0.99        \\
\multicolumn{6}{l}{\textbf{Agent}}                                                       \\
Entropy Coef. Start    & 1      & 1           & 0.5          & 0.5         & 0.2         \\
Entropy Decay Strategy & Linear & Linear      & Linear       & Exponential & Exponential \\
Entropy Decay          & 0.0005 & 0.0007      & 0.00004      & 20000       & 10000       \\
Entropy Steps          & –      & –           & –            & 20000       & 10000       \\
Min. Entropy Coef.     & 0.001  & 0.001       & 0.01         & 0.01        & 0.001       \\
LR Critic              & 8e-4   & 8e-4        & 1e-3         & 1e-3        & 1e-3        \\
LR Actor               & 4e-4   & 4e-4        & 1e-3         & 1e-3        & 5e-4        \\
Hidden Layer size       & 8      & 8           & 16           & 16          & 16          \\
\# Layers              & 2      & 2           & 2            & 2           & 2           \\
\multicolumn{6}{l}{\textbf{Mediator}}                                                    \\
Entropy Coef. Start    & 1      & 1           & 0.5          & 0.5         & 0.2         \\
Entropy Decay Strategy & Linear & Linear      & Linear       & Exponential & Exponential \\
Entropy Decay          & 0.0005 & 0.0007      & 0.00004      & 20000       & 10000       \\
Entropy Steps          & –      & –           & –            & 20000       & 10000       \\
Min. Entropy Coef.     & 0.001  & 0.001       & 0.01         & 0.01        & 0.001       \\
LR Critic              & 1e-3   & 1e-3        & 1e-3         & 1e-3        & 1e-3        \\
LR Actor               & 8e-4   & 8e-4        & 1e-3         & 1e-3        & 5e-4        \\
LR Lambda              & –      & –           & 1e-3         & 1e-3        & 1e-3        \\
\# Hidden Layers       & 8      & 8           & 32           & 16          & 16          \\
\# Layers              & 2      & 2           & 2            & 2           & 2           \\ \bottomrule
\end{tabular}
\end{table*}



\end{document}